\documentclass[a4paper]{PoS}
\usepackage{amssymb}
\usepackage{amsmath}
\usepackage{subfigure}

\title{Cosmogenic gamma-rays and neutrinos constrain UHECR source models}

\ShortTitle{Cosmogenic gamma-rays and neutrinos constrain UHECR source models}

\author{\speaker{Arjen van Vliet}%\thanks{A footnote may follow.}
        \\
       Department of Astrophysics/IMAPP, Radboud University, Nijmegen, The Netherlands\\
       E-mail: \email{a.vanvliet@astro.ru.nl}}

\author{J\"org R. H\"orandel\\
        Department of Astrophysics/IMAPP, Radboud University, Nijmegen, the Netherlands\\
        E-mail: \email{j.horandel@astro.ru.nl}}       

\author{Rafael Alves Batista\\
        Department of Physics - Astrophysics, University of Oxford, Oxford, United Kingdom\\
        E-mail: \email{rafael.alvesbatista@physics.ox.ac.uk}}
        
\abstract{
\textbf{Purpose.} When ultra-high-energy cosmic rays (UHECRs) propagate through the universe they produce secondary neutrinos as well as photons, electrons and positrons (initiating electromagnetic cascades) in different kinds of interactions. These neutrinos and electromagnetic cascades are detected at Earth as isotropic extragalactic fluxes. The level of these fluxes can be predicted and used to constrain UHECR source models.

\textbf{Methods.} The public astrophysical simulation framework CRPropa 3, designed for simulating the propagating extraterrestrial ultra-high energy particles, is ideally suited for this purpose. CRPropa includes all relevant UHECR interactions as well as secondary neutrino and electromagnetic cascade production and propagation. It is designed for high-performance computing and provides the flexibility to scan large parameter ranges of UHECR models. 

\textbf{Results.} The expected cosmogenic neutrino and gamma-ray spectra depend strongly on the evolution with redshift of the UHECR sources and on the chemical composition of UHECRs at injection. The isotropic diffuse gamma-ray background measured by Fermi/LAT is already close to touching upon a model with co-moving source evolution and with the chemical composition, spectral index and maximum acceleration energy optimized to provide the best fit to the UHECR spectrum and composition measured by the Pierre Auger Collaboration. Additionally, the detectable fraction of protons present at the highest energies in UHECRs is shown as a function of the evolution of UHECR sources for a range of sensitivities of neutrino detectors at an energy of $\sim1$~EeV.

\textbf{Conclusions.} Neutrino and gamma-ray measurements are starting to constrain realistic UHECR models. Current and future neutrino experiments with sensitivities in the range of $\sim 10^{-8}$ -- $10^{-10}$ GeV cm$^{-2}$ s$^{-1}$ sr$^{-1}$ for the single-flavor neutrino flux at $\sim1$~EeV will be able to significantly constrain the proton fraction for realistic source evolution models.
}

\FullConference{35th International Cosmic Ray Conference --- ICRC2017\\
		10--20 July, 2017\\
		Bexco, Busan, Korea}

\begin{document}

\section{Introduction}\label{intro}

During the propagation of ultra-high-energy cosmic rays (UHECRs) through the universe they can interact with photons from the cosmic microwave background (CMB) and the extragalactic background light (EBL). In these interactions secondary neutrinos as well as photons, electrons and positrons can be created. These photons, electrons and positrons may again interact with the ambient photon backgrounds and initiate electromagnetic cascades. The neutrinos from these interactions and the gamma-rays created in these electromagnetic cascades can then arrive at Earth as isotropic extragalactic cosmogenic fluxes. The level of these fluxes can be predicted with simulations of the propagation of UHECRs and depends on the properties of UHECR sources. Therefore, in this way measurements on UHECRs, neutrinos and gamma-rays can be used together to constrain models of UHECR sources.

The energy spectrum of UHECRs has been measured with unparalleled precision by the Pierre Auger Observatory (Auger)~\cite{Aab:2015bza,ThePierreAuger:2015rha} and the Telescope Array (TA)~\cite{Jui:2015tac}. The mass composition results obtained by Auger indicate an increasingly heavier mass for $E \gtrsim10^{9.3}$~GeV~\cite{Porcelli:2015pac}. In the same energy range, when following the argumentation in Ref.~\cite{Fujii:2015tac}, the TA data are consistent with a pure proton composition, while being consistent with Auger data as well~\cite{Unger:2015ptc}. The pure proton interpretation is, however, disfavoured by the analysis done by Auger in Ref.~\cite{Aab:2016htd}. Since the Auger data is more precise in terms of statistical and systematic uncertainty, we will focus on that data set in the following.

IceCube has measured an astrophysical neutrino flux in the energy range between 60~TeV and 3~PeV~\cite{Aartsen:2015zva} and has provided upper limits up to an energy of 100 EeV~\cite{Aartsen:2016ngq}. Additionally, Auger has supplied upper limits on the neutrino flux above 100~PeV~\cite{Aab:2015kma} and ANITA-II above 1~EeV~\cite{Gorham:2010kv}. Cosmogenic neutrinos are not the only possible source of astrophysical neutrinos in this energy range, they could be created inside the sources of UHECRs as well. Therefore the measured flux by IceCube should also be considered as an upper bound. The same holds in the gamma-ray case for the measurements of the isotropic diffuse gamma-ray background (IGRB) by Fermi/LAT~\cite{Ackermann:2014usa}. In fact in Refs.~\cite{Lisanti:2016jub,TheFermi-LAT:2015ykq} has been shown that between $60\%$ to $100\%$ of the IGRB consists of unresolved point sources. The Fermi/LAT IGRB can therefore be shifted down by a factor of 0.4 at least to provide an upper limit for the cosmogenic gamma-ray flux. 

In recent years several investigations have been done by different groups to constrain the sources of UHECRs using cosmogenic fluxes of neutrinos and/or photons, disfavoring pure proton models with a strong source evolution (see e.g. Refs.~\cite{Aloisio:2015ega,Heinze:2015hhp,Gavish:2016tfl,Berezinsky:2016jys,Supanitsky:2016gke,Globus:2017ehu}). Here we show how these results depend on the different simulation parameters in Sec.~\ref{Deps}, what can be expected for a model that fits both the UHECR spectrum and composition measurements of Auger in Sec.~\ref{BestFit} and, focusing on the neutrino flux at $\sim 1$~EeV, which fraction of protons at the highest energies in UHECRs is detectable in this way (Sec.~\ref{NuAt1EeV}). 

\section{Simulation setup}
\label{SimSetup}

In this work we use CRPropa 3~\cite{Batista:2016yrx} to simulate the propagation and production of UHECRs and their secondary gamma-rays and neutrinos. CRPropa is a public simulation framework for Monte Carlo propagation of UHECRs and their secondaries including all relevant interactions, namely: photopion production, pair production, adiabatic losses, as well as photodisintegration and nuclear decay (in the case of nuclei). The propagation of electromagnetic cascades is simulated using the specialized code DINT~\cite{Lee:1996fp}, which is part of CRPropa. DINT solves the one-dimensional transport equations for electromagnetic cascades initiated by photons or leptons. It includes all relevant energy losses namely single, double, and triplet pair production, inverse Compton scattering and synchrotron emission. 

All the simulations done here are one-dimensional assuming a homogeneous distribution of identical sources. The cosmic rays are injected by sources following a spectrum of
\begin{equation}\label{eq:PowerLawInjection}
  \frac{\text{d}N}{\text{d}E} \propto \left( \dfrac{E}{E_0} \right)^{-\alpha} \exp\left( -\dfrac{E}{Z R_{\text{cut}}} \right) ,
\end{equation}
wherein $E$ is the energy of the particles, $E_0$ an arbitrary normalization energy, $\alpha$ the spectral index, $Z$ the charge of the primary cosmic ray, and $R_{\text{cut}} \equiv E_{\text{cut}}/Z$ is the cutoff rigidity. A minimum energy of $E_{\text{min}} = 0.1$~EeV is assumed for the cosmic rays at the source. We use the EBL model by Gilmore {\it et al.}~\cite{Gilmore:2011ks}. The sources evolve as $(1 + z)^m$, where $m$ is the evolution parameter and $z$ the redshift of the source, up to a maximum redshift of $z_{\text{max}}=6$. The resulting UHECR flux is normalized to the measured spectrum of Auger~\cite{Aab:2015bza} at $E = 10^{9.85}$~GeV and the neutrino and gamma-ray results are normalized accordingly. 

\section{Simulation parameter dependencies}
\label{Deps}

We investigate how the simulated UHECR, neutrino and gamma-ray fluxes depend on the different source parameters in the simulation. First, we take as a reference scenario a pure proton case with $R_{\text{cut}} = 200$~EV, $\alpha = 2.5$ and $m=0$. In the next sections the effects of changing $R_{\text{cut}}$ and $\alpha$ are investigated individually; the role of the source evolution $m$ is discussed in Ref.~\cite{vanVliet:2016dyx}.

The values of these parameters are not optimized to fit the cosmic-ray spectrum perfectly as the systematic uncertainty of the energy spectrum measurements is much larger than the statistical uncertainty of both Auger and TA. Furthermore the assumption made here, of a homogeneous distribution of identical sources with a power-law injection spectrum with exponential cutoff, may not correctly depict the real situation. 
Fitting a pure proton model to the TA spectrum would lead to a strong source evolution and a relatively hard injection spectrum and would be strongly constrained by both the Fermi/LAT IGRB measurement and the IceCube data~\cite{Heinze:2015hhp,Supanitsky:2016gke}.

The results of the reference model are shown in Figs.~\ref{fig:Emax}, \ref{fig:Alpha}, \ref{fig:Mass} and \ref{fig:BestFit} as solid lines. The UHECR spectra are compared with measurements from Auger~\cite{Aab:2015bza} and TA~\cite{Jui:2015tac}; the single-flavor neutrino spectra are shown together with the IceCube data~\cite{Aartsen:2015zva} and upper limits from IceCube~\cite{Aartsen:2016ngq}, Auger~\cite{Aab:2015kma} and ANITA-II~\cite{Gorham:2010kv}. The cosmogenic photons are displayed with the Fermi/LAT IGRB~\cite{Ackermann:2014usa} and this same IGRB multiplied by a factor of 0.4 to account for the fact that more than $60\%$ of the IGRB is produced by unresolved point sources~\cite{Lisanti:2016jub,TheFermi-LAT:2015ykq}. 

From Fig.~\ref{fig:EmaxCRs} it can be seen that the reference model ($R_{\text{cut}} = 200$~EV) satisfactorily reproduces the measured Auger spectrum for $E \gtrsim 2\times10^{9}$~GeV. Note that the lower energy range ($E \lesssim 2\times10^{9}$~GeV) might also be the regime where a Galactic contribution to the cosmic-ray spectrum starts playing a role, which could possibly compensate for an underprediction of the cosmic-ray spectrum. Fig.~\ref{fig:EmaxNeutrinos} shows that the simulated cosmogenic neutrino flux remains below the measurements and limits by IceCube, Auger and ANITA-II for the whole energy range considered here. The photon case, however, is already in slight tension with the highest energy bins of the IGRB x 0.4, as can be seen from Fig.~\ref{fig:EmaxPhotons}.

\subsection{Maximum source energy}
\label{EcutDep}

\begin{figure}
  \centering
  \subfigure[Cosmic rays]{
       \includegraphics[width=.315\textwidth]{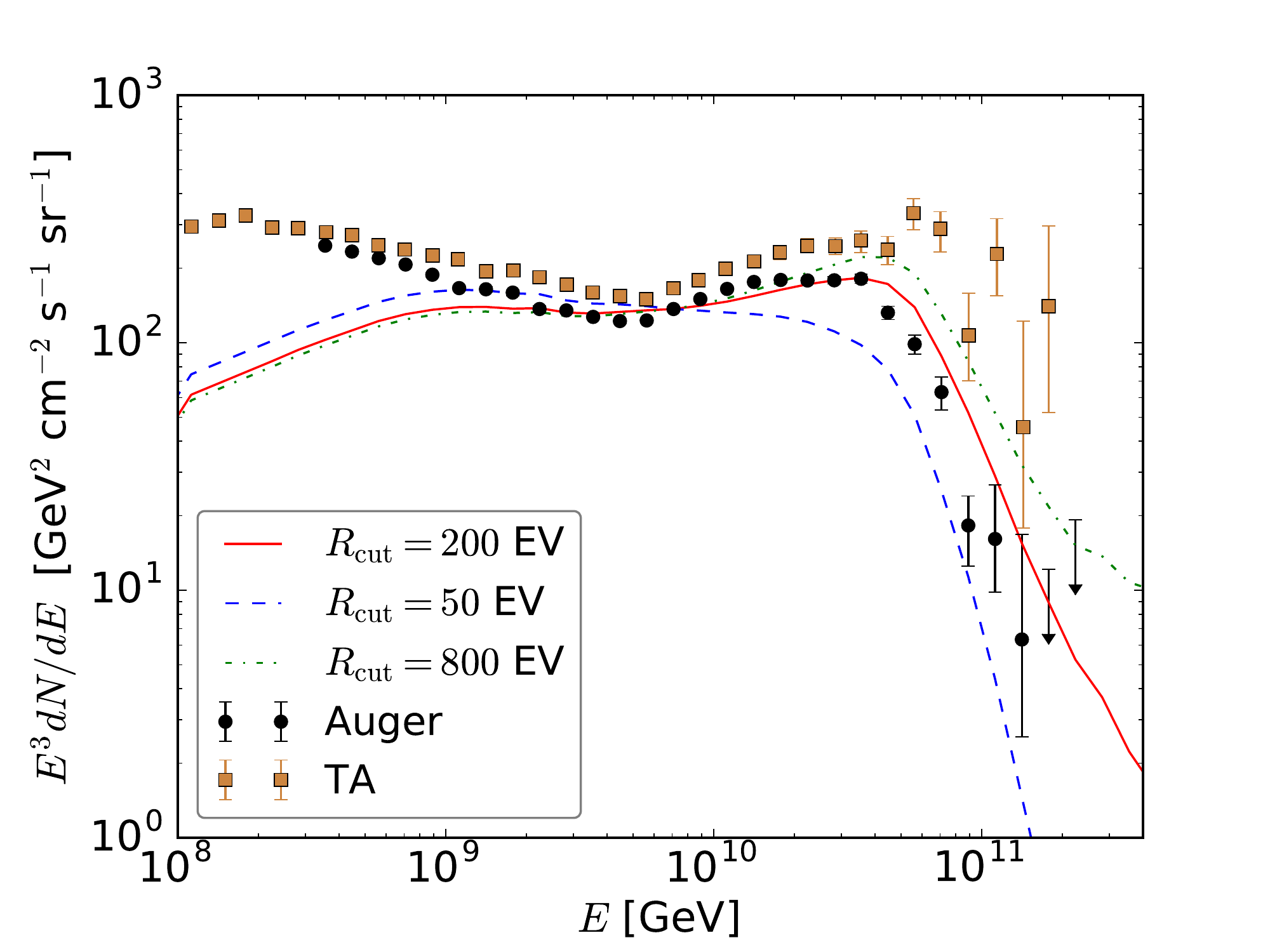}\label{fig:EmaxCRs}
     }
     \subfigure[Neutrinos]{
       \includegraphics[width=.315\textwidth]{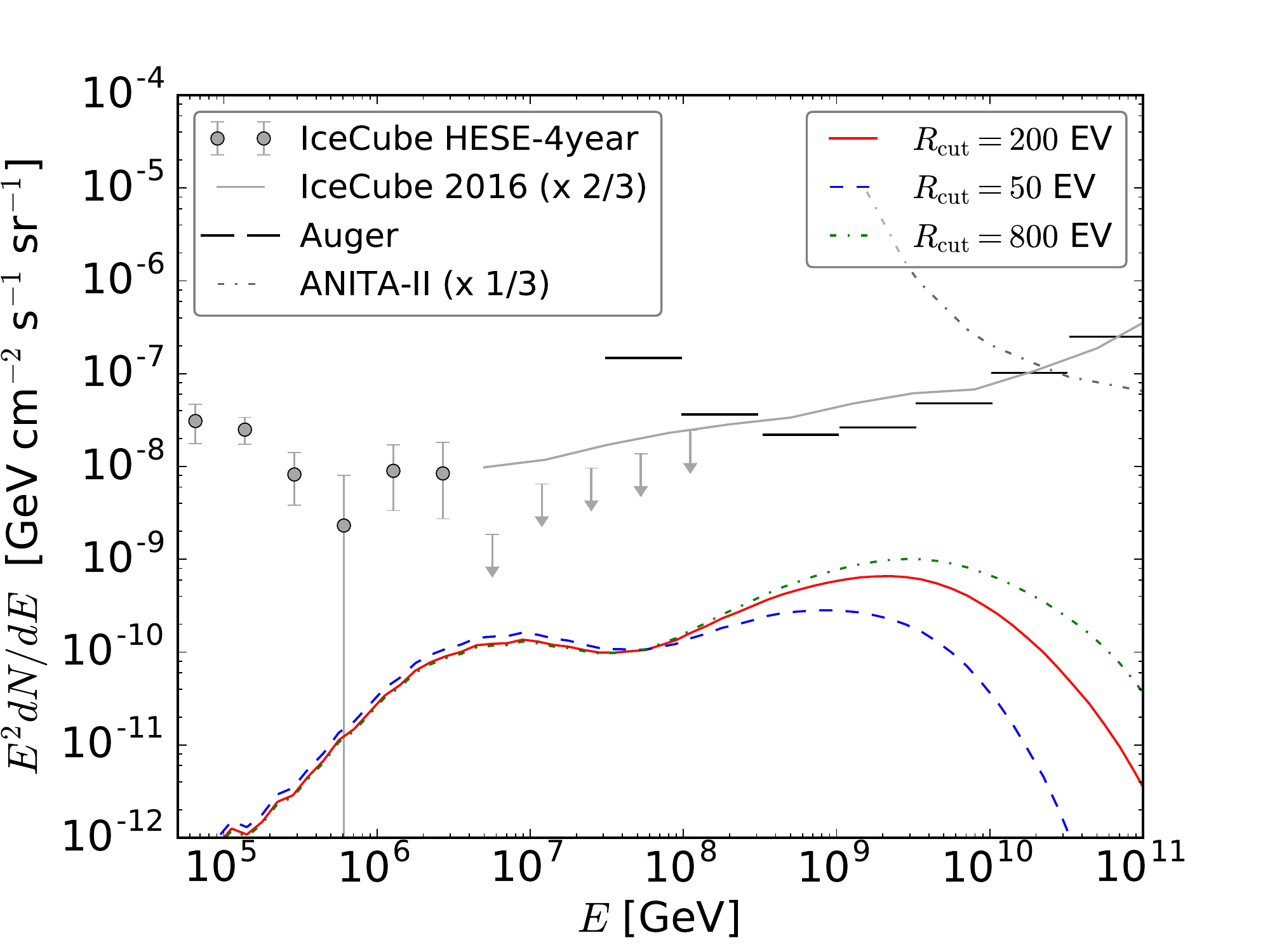}\label{fig:EmaxNeutrinos}
     }
     \subfigure[Photons]{
       \includegraphics[width=.315\textwidth]{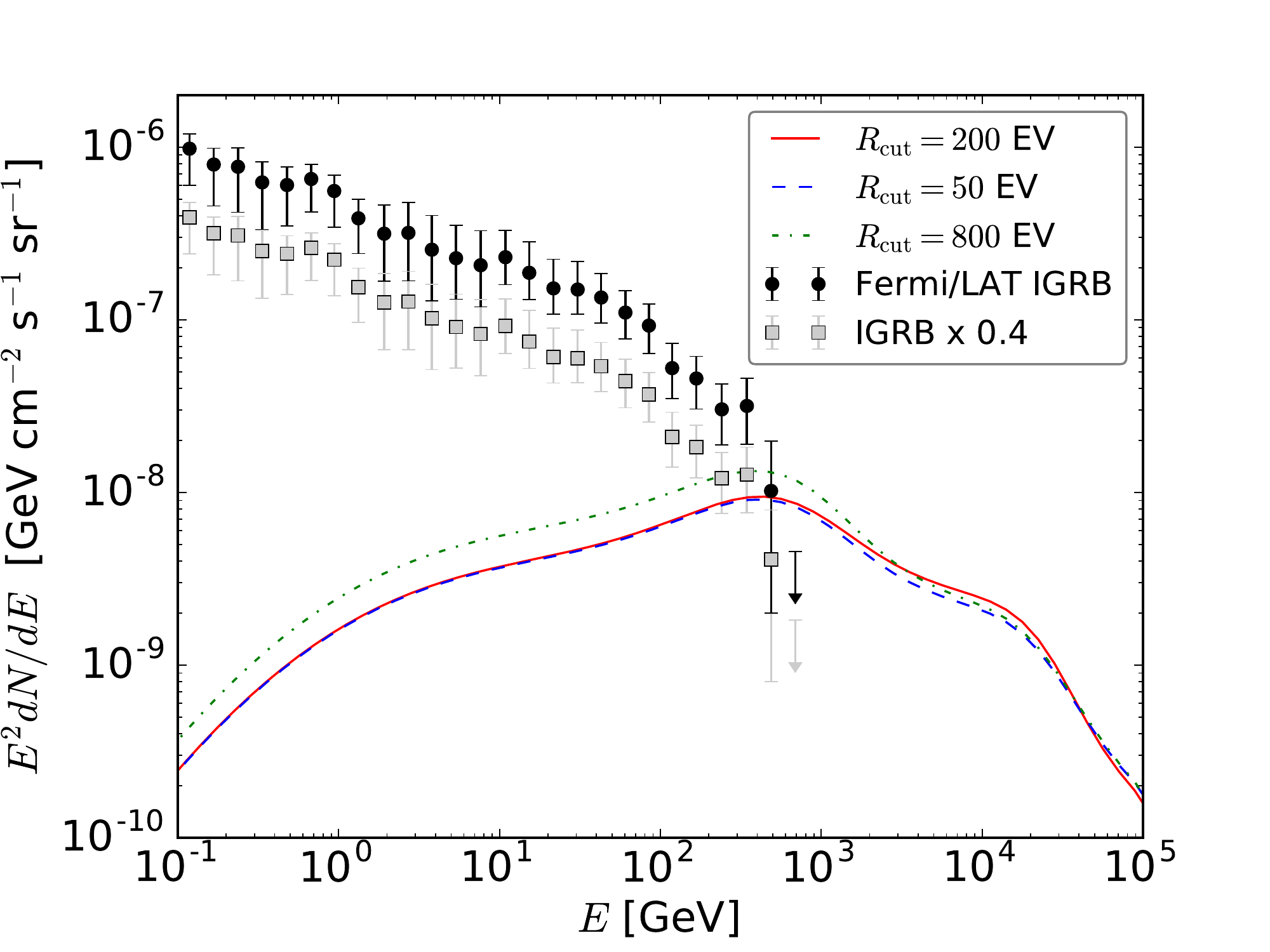}\label{fig:EmaxPhotons}
     }
   \caption{(a) Simulated UHECR spectrum for pure-proton models with $\alpha = 2.5$ and $R_{\text{cut}} = $ 50, 200 and 800~EV. (b) Single-flavor neutrino spectra (c) Photon spectra. See text for further details.}
   \label{fig:Emax}
 \end{figure}

The expected flux of cosmogenic neutrinos and photons is affected by the maximum energy attainable by cosmic accelerators ($R_{\text{cut}}$). In Fig.~\ref{fig:Emax} the results are given for varying $R_{\text{cut}}$ between 50~EV (dashed lines) and 800~EV (dashed-dotted lines) keeping all other parameters fixed.

From Fig.~\ref{fig:EmaxCRs} it can be seen that the value of $R_{\text{cut}}$ has a significant effect on the high-energy region of the UHECR spectrum. However, increasing $R_{\text{cut}}$ to values even higher than 800~EV (while keeping all other parameters the same) would not significantly affect the UHECR spectrum, since in this case the dominant process behind the cutoff in the UHECR spectrum is the Greisen-Zatsepin-Kuzmin (GZK) effect~\cite{Greisen:1966jv,Zatsepin:1966jv}, not the maximum source energy. Decreasing $R_{\text{cut}}$ to values below 50~EV, for this pure proton case, would lessen the agreement with the measured data to a point that can no longer be considered realistic. Fig.~\ref{fig:EmaxNeutrinos} shows that $R_{\text{cut}}$ only affects the expected neutrino spectrum for $E \gtrsim 10^{8}$~GeV and remains below the limits for the whole energy range. Fig.~\ref{fig:EmaxPhotons} indicates that the expected gamma-ray flux increases only slightly for $R_{\text{cut}}>200$~EV and remains basically the same for $R_{\text{cut}}<200$~EV. 

\subsection{Spectral index}
\label{AlphaDep}

\begin{figure}
  \centering
  \subfigure[Cosmic rays]{
       \includegraphics[width=.315\textwidth]{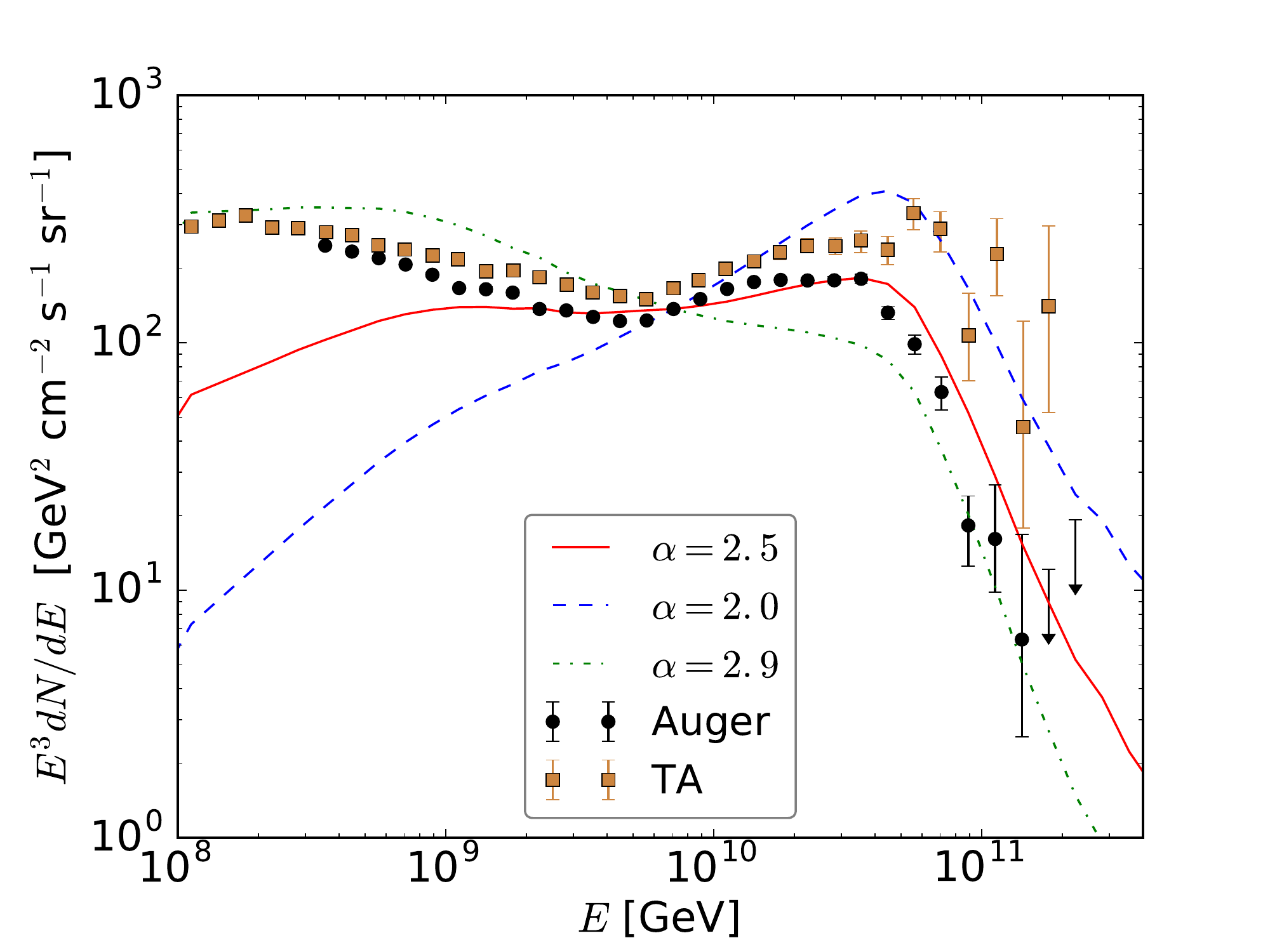}\label{fig:AlphaCRs}
     }
     \subfigure[Neutrinos]{
       \includegraphics[width=.315\textwidth]{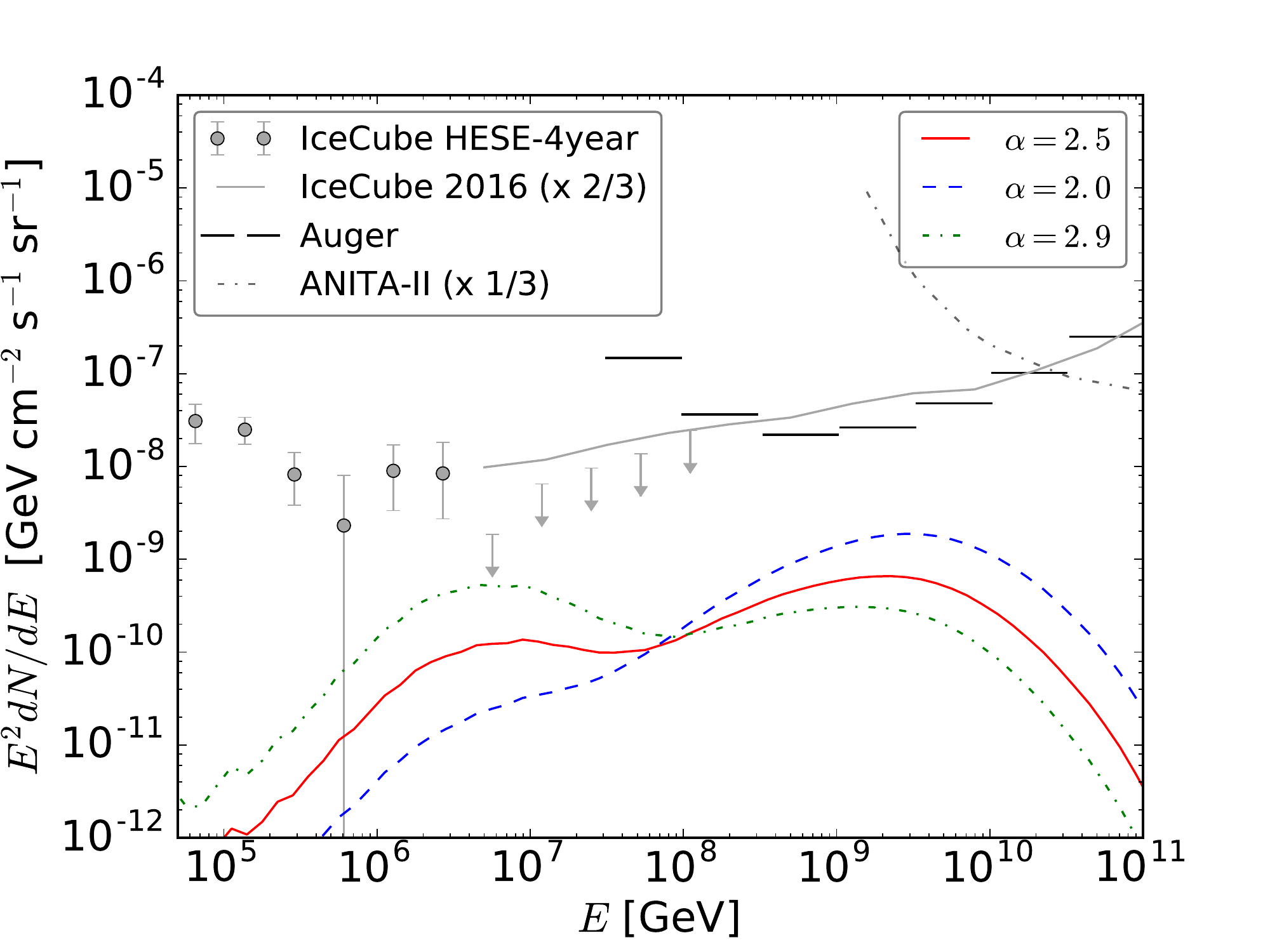}\label{fig:AlphaNeutrinos}
     }
     \subfigure[Photons]{
       \includegraphics[width=.315\textwidth]{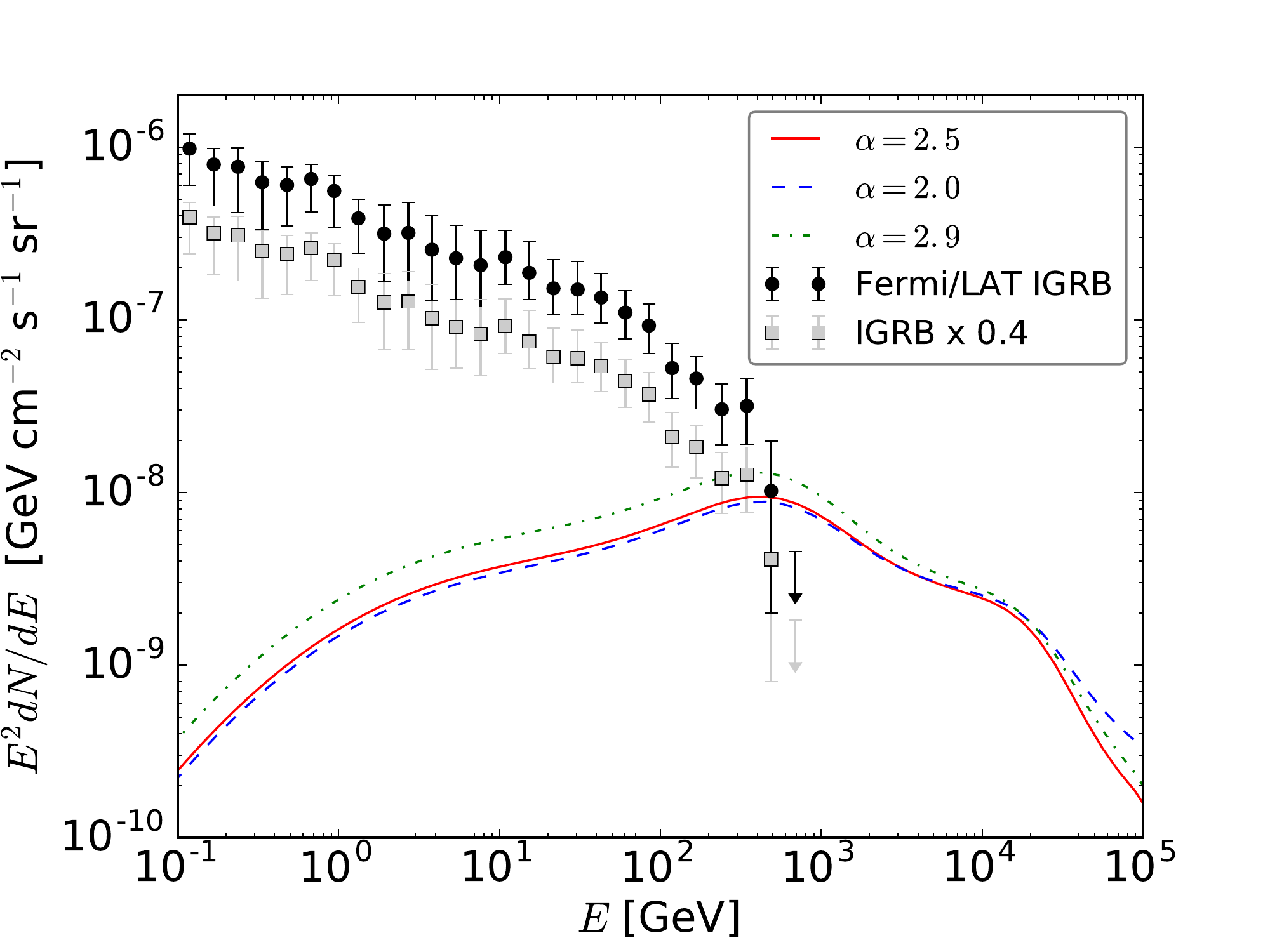}\label{fig:AlphaPhotons}
     }
   \caption{(a) Simulated UHECR spectrum for pure-proton models with $R_{\text{cut}} = 200$~EV and $\alpha = 2.0$, 2.5 and 2.9. (b) Single-flavor neutrino spectra. (c) Photon spectra. See text for further details.}
   \label{fig:Alpha}
 \end{figure}

The value of the spectral index $\alpha$ of the power-law injection at the UHECR sources can also affect the expected fluxes. Here we investigate the range $2.0 \leq \alpha \leq 2.9$ whilst keeping all other parameters of the reference scenario fixed. These values of $\alpha$ are in reasonable agreement with the measured UHECR spectrum and encompass the typical spectral indices expected from Fermi acceleration ($\alpha \simeq 2.0 - 2.4$). In this scenario a low value of $\alpha$ within this range would deem the ankle as the spectral feature associated with the transition between extragalactic and galactic cosmic rays; conversely, a high value of $\alpha$ for would imply that such transition occurs at lower energies. Fig.~\ref{fig:Alpha} shows the results for the cosmic-ray, neutrino and photon spectra. 

Fig.~\ref{fig:AlphaNeutrinos} shows that $\alpha$ significantly affects both the neutrino and the UHECR spectrum in a similar way, but its impact on the gamma-ray spectrum (Fig.~\ref{fig:AlphaPhotons})  is small in this energy range.

\subsection{Composition}
\label{MassDep}

\begin{figure}
   \centering
   \subfigure[Cosmic rays]{
       \includegraphics[width=.315\textwidth]{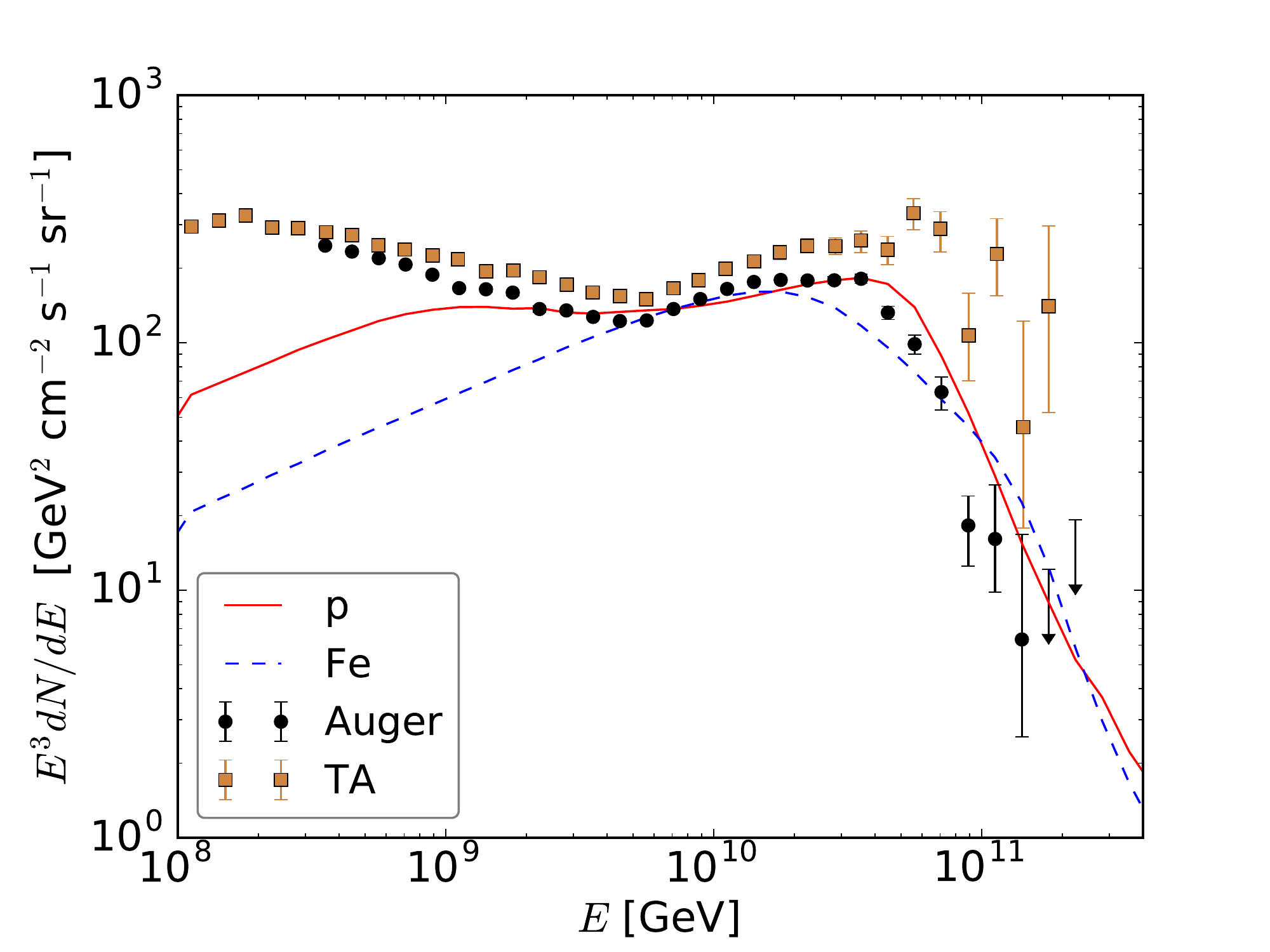}\label{fig:IronCRs}
     }
     \subfigure[Neutrinos]{
       \includegraphics[width=.315\textwidth]{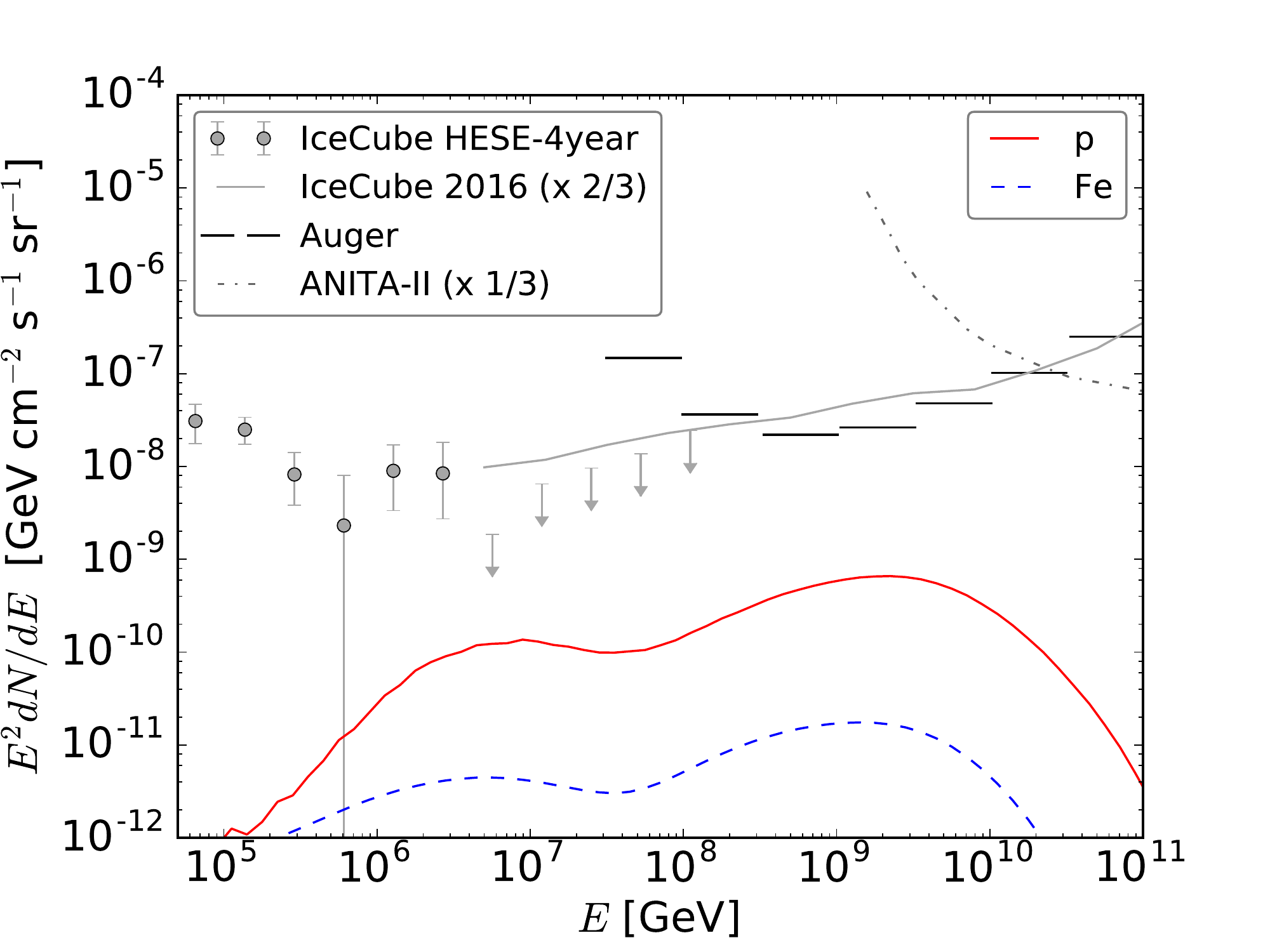}\label{fig:IronNeutrinos}
     }
     \subfigure[Photons]{
       \includegraphics[width=.315\textwidth]{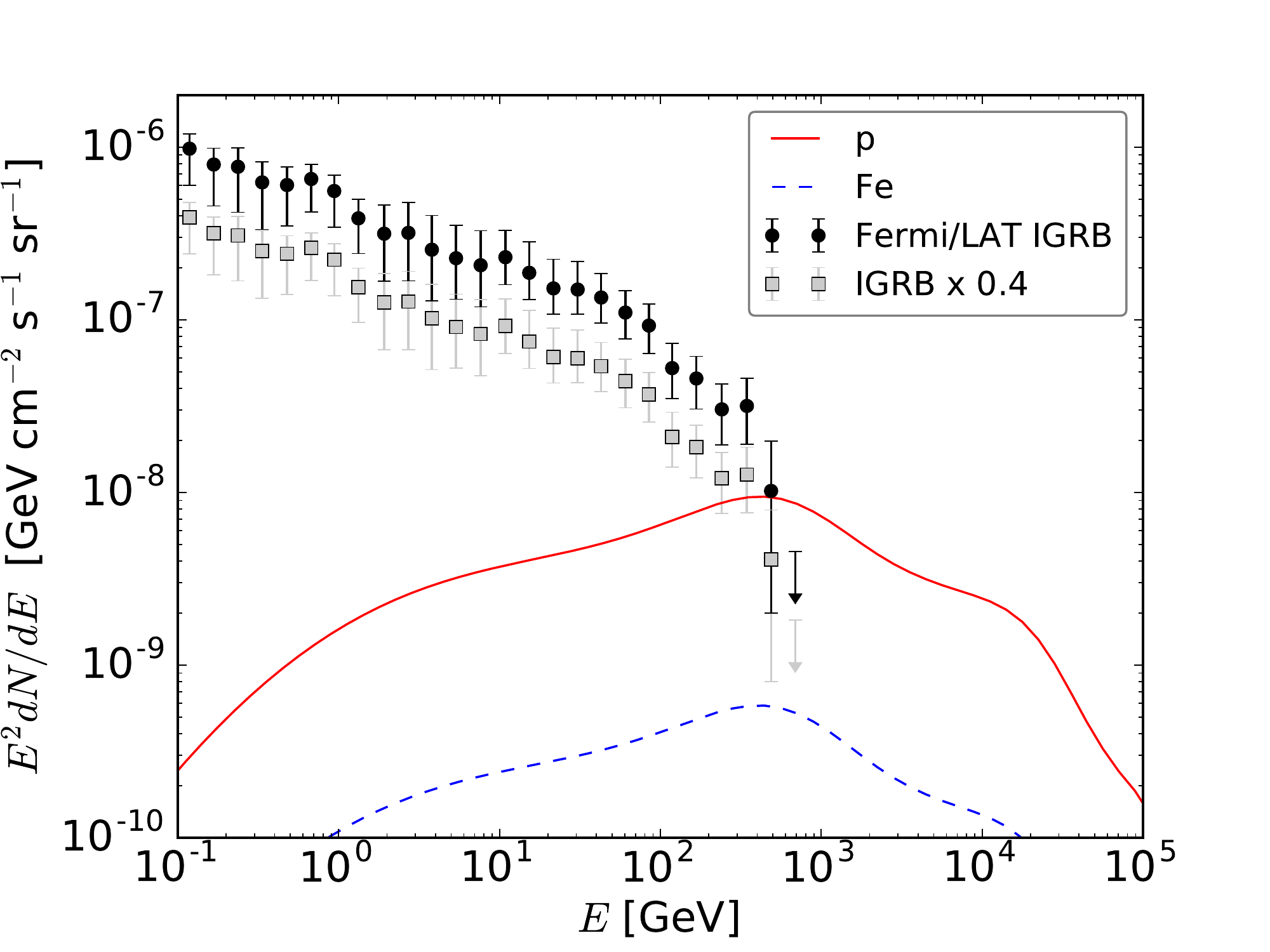}\label{fig:IronPhotons}
     }
   \caption{(a) Simulated UHECR spectrum for pure-proton and pure-iron injection, with $\alpha = 2.5$ and $R_{\text{cut}} = $ 200~EV. (b) Single-flavor neutrino spectra. (c) Photon spectra. See text for further details.}
   \label{fig:Mass}
 \end{figure}

Another factor that can play a role in the determination of the cosmogenic neutrino and photon fluxes is the initial mass composition of the cosmic rays. In the case of a pure proton composition the dominant interactions are photopion production and pair production during which several leptons and/or photons are created. When starting with heavy nuclei the main interaction is photodisintegration in which no leptons and far less photons are created. To see how this affects the expected neutrino and photon spectra we compare the pure proton reference scenario with a case where only iron nuclei are injected at the sources while keeping all other parameters of the simulation the same. See Fig.~\ref{fig:Mass} for the results.

When switching to heavy nuclei instead of protons it is no longer possible to explain the ankle in the cosmic-ray spectrum as due to effects of interactions during the propagation. From Fig.~\ref{fig:IronCRs} can therefore be seen that the iron case starts deviating from the measured spectrum at the ankle. Figs.~\ref{fig:IronNeutrinos} and \ref{fig:IronPhotons} show that both the cosmogenic neutrino and photon fluxes are strongly reduced when starting with iron nuclei instead of with protons.

\section{Combined-fit model}
\label{BestFit}

\begin{figure}
  \centering
     \subfigure[Neutrinos]{
       \includegraphics[width=.315\textwidth]{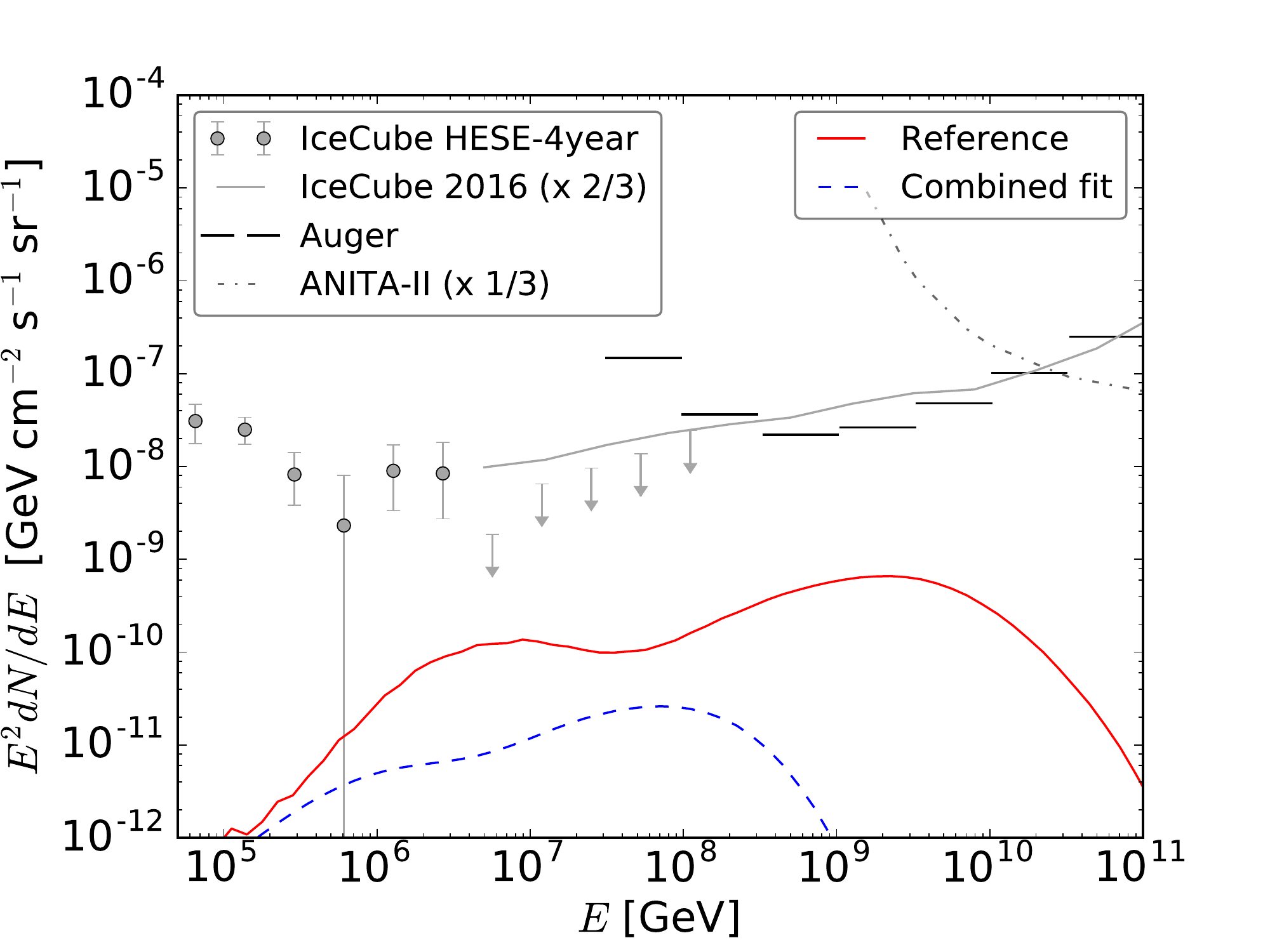}\label{fig:AugerFitNeutrinos}
     }
     \subfigure[Photons]{
       \includegraphics[width=.315\textwidth]{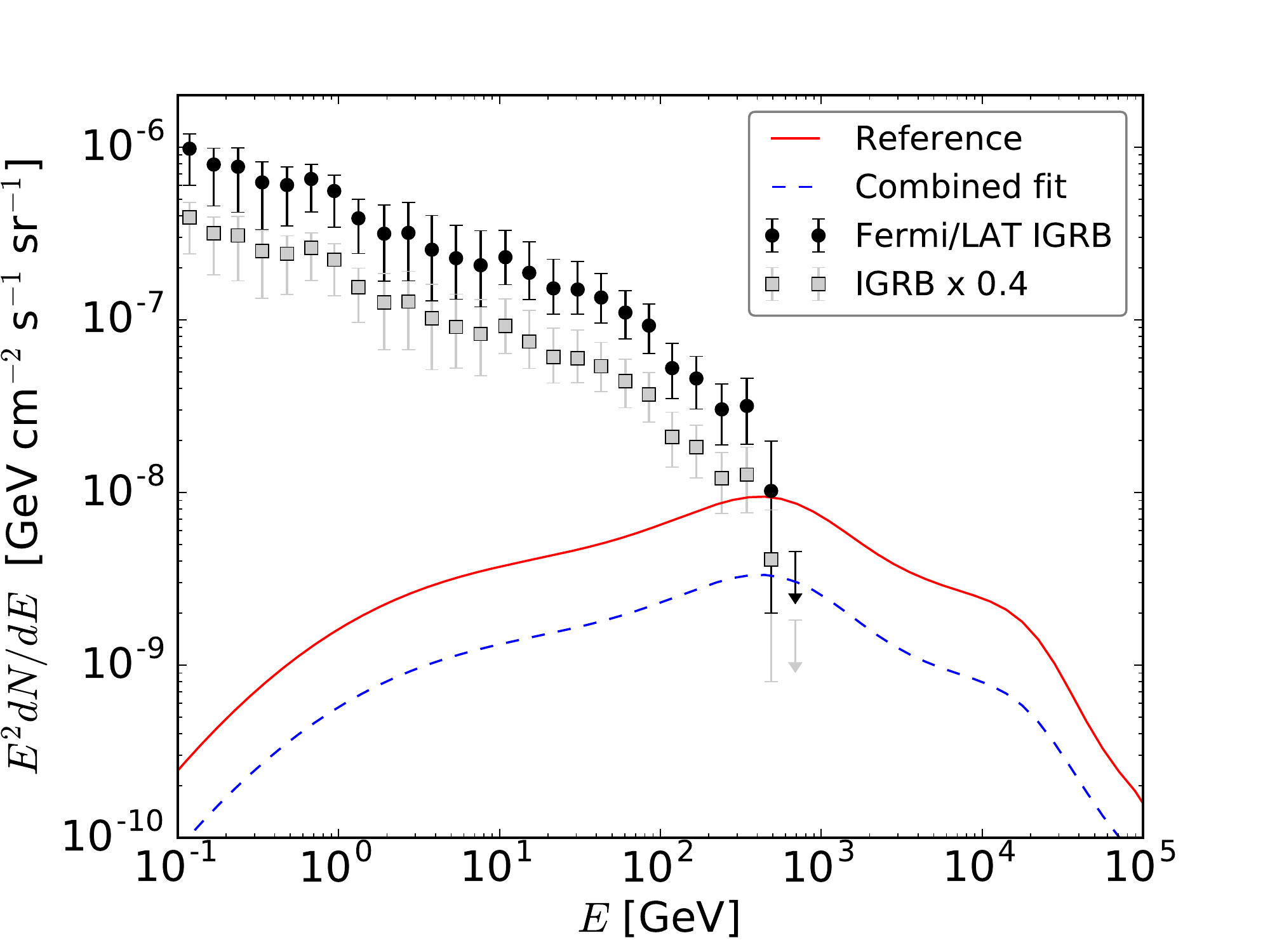}\label{fig:AugerFitPhotons}
     }
   \caption{Simulated neutrino (a) and gamma-ray (b) fluxes for Auger's CTG best-fit model with $\alpha = 0.87$, $R_{\text{cut}} = 10^{0.62}$~EV, $f_{\text{N}} = 88\%$ and $f_{\text{Si}} = 12\%$. The reference scenario is also shown. See text for further details.}
   \label{fig:BestFit}
 \end{figure}

Auger has recently performed a combined spectrum-composition fit of their data above an energy of $5\cdot10^{9}$~GeV~\cite{Aab:2016zth}. They assume that identical sources are accelerating particles through a rigidity-dependent mechanism, and that these sources are distributed uniformly in the co-moving volume. Their results favor sources with a relatively low maximum energy, a hard injection spectrum and a heavy composition. We consider one of their models labeled `CTG'. They employ this power-law injection with exponential cutoff of UHECRs: 
\begin{equation}\label{eq:BFPowerLawInjection}
	\frac{\text{d}N}{\text{d}E} \propto \left\{
	\begin{array}{l l}
 	(E/E_0)^{-\alpha} & E \leq E_{\text{cut}} \\
 	(E/E_0)^{-\alpha}\exp(1-E/(ZR_{\text{cut}})) & E > ZR_{\text{cut}}
	\end{array}
	\right.
\end{equation}
They find two main local minima whose total deviances differ by an insignificant amount. Of these two minima, we take here the one with $\alpha$ closest to standard Fermi acceleration. The parameters for this minimum are: $\alpha = 0.87$, $R_{\text{cut}} = 10^{0.62}$~EV, $f_{\text{N}} = 88\%$ and $f_{\text{Si}} = 12\%$, with $f_{\text{N}}$ and $f_{\text{Si}}$ the elemental fractions at injection at $E_0 = 1$~EeV. The resulting cosmogenic neutrino and photon spectra are given in Fig.~\ref{fig:BestFit}, compared with the pure proton reference scenario discussed in Sec.~\ref{Deps}. 

The neutrino peak at $E > 10^9$~GeV completely disappears in this best-fit scenario (see Fig.~\ref{fig:AugerFitNeutrinos}) due to the relatively low maximum energy. Due to the heavy composition both the cosmogenic neutrino and photon fluxes are reduced with respect to the reference scenario. Still, with a few more years of data and taking into account that most of the IGRB comes from unresolved point sources, Fermi/LAT might be able to touch upon this cosmogenic gamma-ray flux. 

\section{Neutrinos at 1~EeV}
\label{NuAt1EeV}

The best-fit model discussed in Sec.~\ref{BestFit} assumes rigidity-dependent cutoffs at the sources with a relatively low maximum rigidity. In that way there are no protons left at all at the highest energies. The composition data of Auger, however, does allow for a non-negligible proton percentage at the highest energies. If such a proton component exists and could be distinguished from the other cosmic rays it could be used to do cosmic-ray astronomy~\cite{AlvesBatista:2017vob}, as the deflections in magnetic fields of such a proton component would be much smaller than the deflections of heavy nuclei of the same energy. Also, if such a proton component exists, it would create much more neutrinos at EeV energies than are present in the best-fit model discussed in Sec.~\ref{BestFit}.

At $E =$ 1~EeV the cosmogenic neutrino flux of such a proton component does not depend much on either $\alpha$, or $R_{\text{cut}}$ (as long as it is not too low) or the specific EBL model, as shown in Fig.~\ref{NuAt1EeV} for $\alpha = 2.0$, 2.5 and 2.9, $R_{\text{cut}} = 100$, $10^{3}$ and $10^{5}$~EV, and $m=3$ for $z<1.5$ and $m=0$ otherwise, up to a maximum redshift of $z_{\text{max}} = 4$. These simulations are done using the EBL model of Franceschini {\it et al.}~\cite{Franceschini:2008tp} and are normalized to the Auger spectrum at $E = 10^{10.55}$~GeV. 

\begin{figure}
  \centering
  \subfigure[Cosmic rays]{
       \includegraphics[width=.315\textwidth]{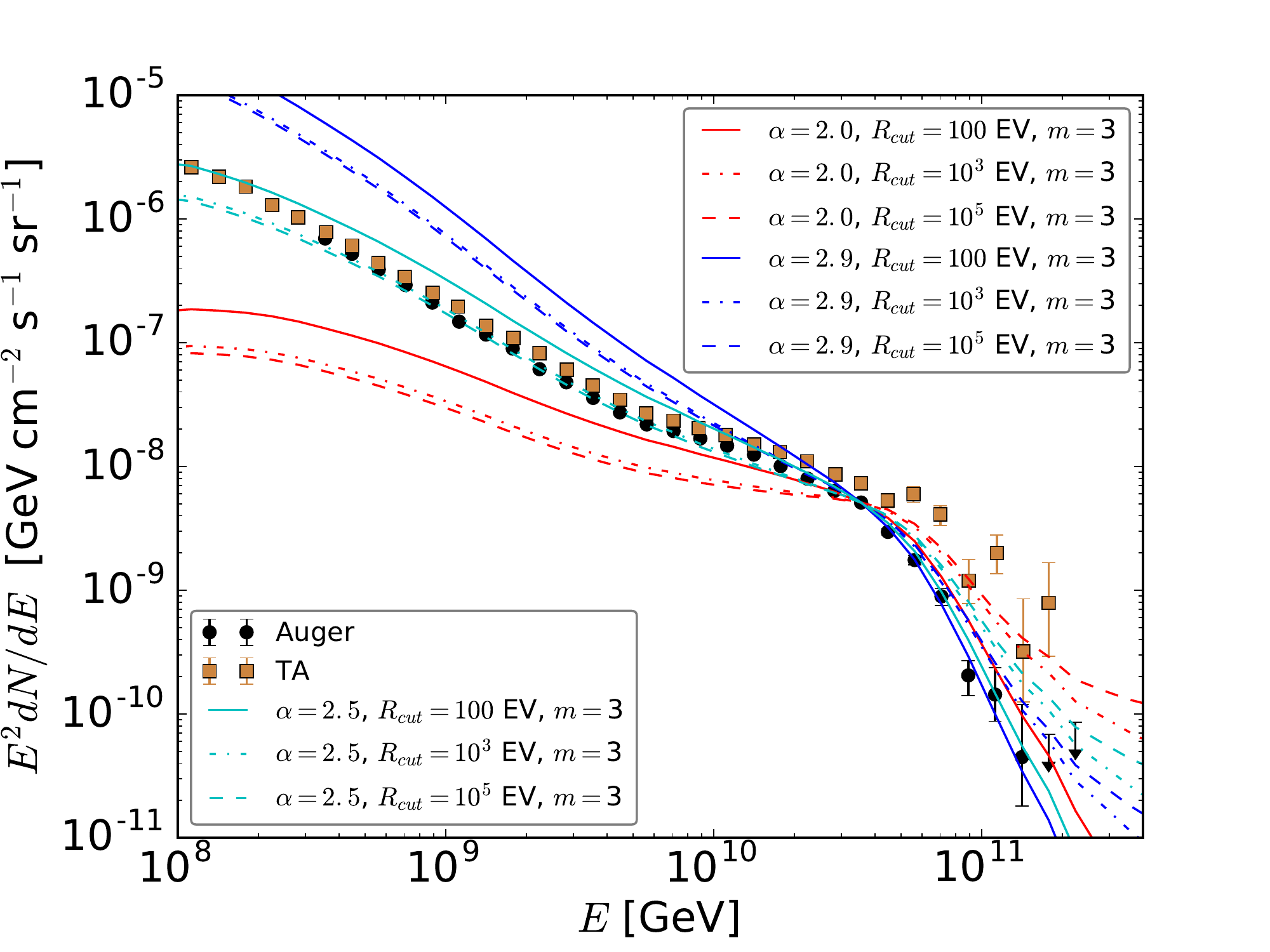}\label{fig:NuAt1EeVCRs}
     }
     \subfigure[Neutrinos]{
       \includegraphics[width=.315\textwidth]{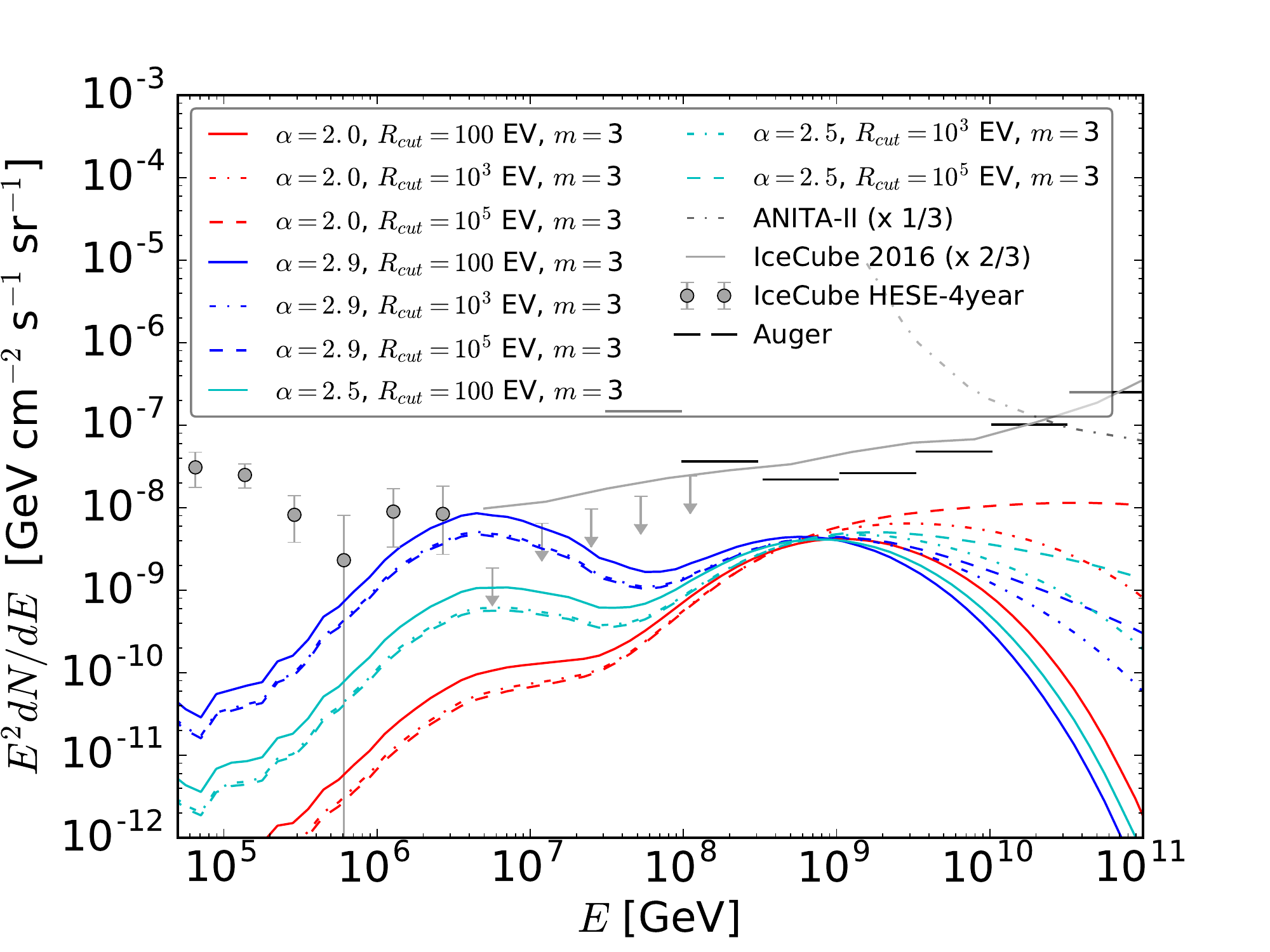}\label{fig:NuAt1EeVneutrinos}
     }
     \subfigure[Detectable fraction of protons]{
       \includegraphics[width=.315\textwidth]{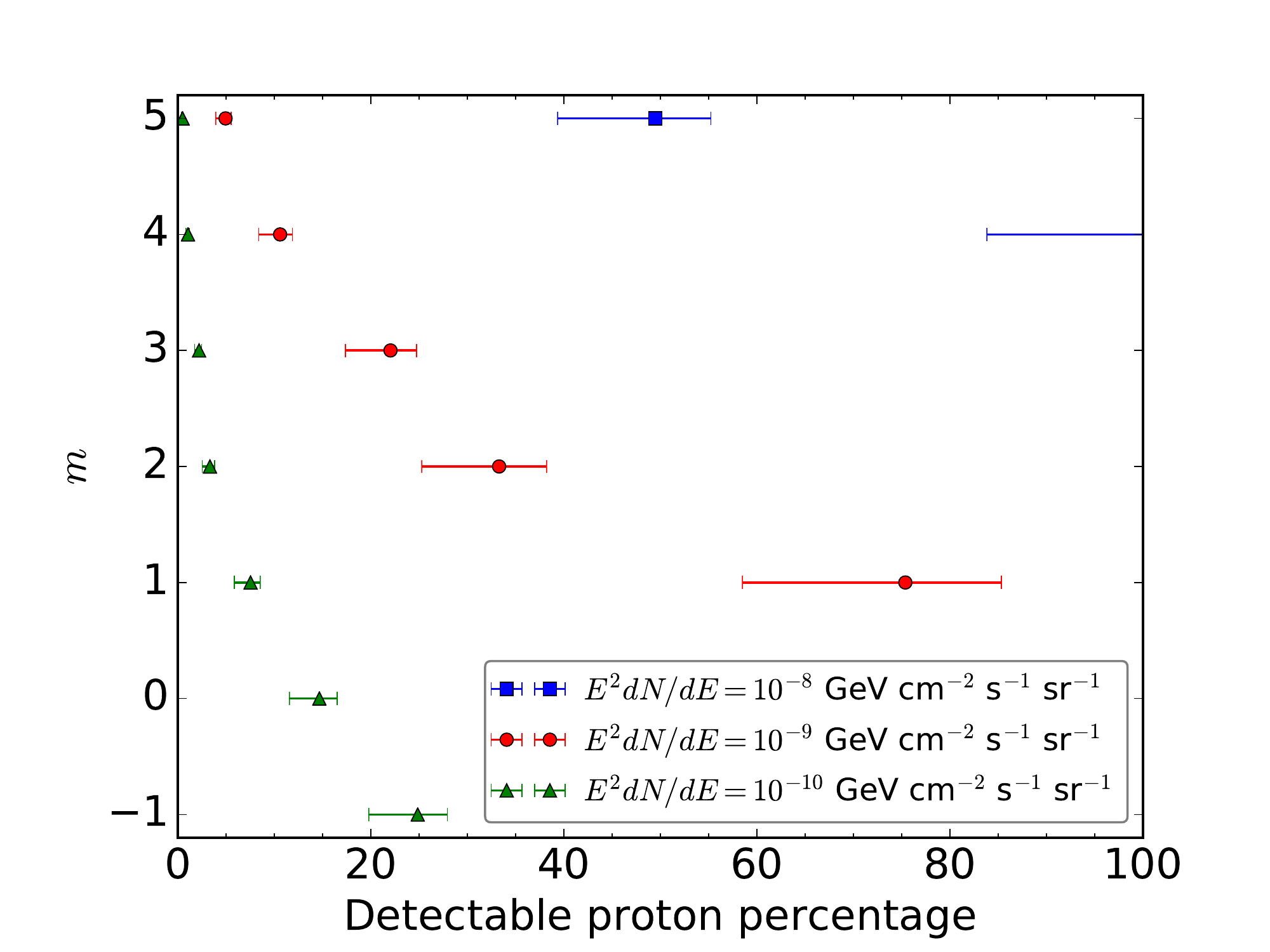}\label{fig:NuAt1EeVpPercent}
     }
   \caption{ (a) Simulated cosmic-ray spectra normalized to the Auger flux~\cite{Aab:2015bza} (circles) at $E = 10^{10.55}$~GeV for $\alpha = 2.0$, 2.5 and 2.9, $R_{\text{cut}} = 100$, $10^{3}$ and $10^{5}$~EV and $m=3$. (b) Corresponding single-flavor neutrino spectra. (c) Detectable proton fraction at the highest energies for neutrino experiments with the indicated sensitivities. The error bars show the spread in neutrino flux due to different values of $\alpha$ and $R_{\text{cut}}$.}
   \label{fig:NuAt1EeV}
 \end{figure}

From Fig.~\ref{fig:NuAt1EeVneutrinos} it can be seen that, for this setup, the spread in cosmogenic neutrino fluxes at $E \approx 1$~EeV is relatively small so that the flux level at this energy does not depend much on $\alpha$ or $R_{\text{cut}}$. The expected cosmogenic neutrino flux will strongly depend on the source evolution parameter $m$, as shown in Ref.~\cite{vanVliet:2016dyx}. In Fig.~\ref{fig:NuAt1EeVCRs} and \ref{fig:NuAt1EeVneutrinos} the simulation results are shown for a $100\%$ proton component at the highest energies. However, because protons produce much more neutrinos than heavier nuclei, as shown in Sec.~\ref{MassDep}, the contribution of heavier nuclei to the cosmogenic neutrino flux at 1~EeV can be safely neglected in this case. In that way we can just scale the total cosmic-ray and neutrino flux down by a certain factor to estimate the cosmogenic neutrino flux for a given proton fraction. The results of this exercise are shown in Fig.~\ref{fig:NuAt1EeVpPercent} and give us the detectable percentage of protons present at the highest energies for certain detector sensitivities of the neutrino flux at 1~EeV, as a function of the source evolution parameter $m$. The error bars in Fig.~\ref{fig:NuAt1EeVpPercent} show the spread in neutrino flux due to different values of $\alpha$ and $R_{\text{cut}}$.

\section{Conclusions}
\label{Conclusions}

We have shown how the expected cosmogenic neutrino and photon spectra depend on the maximum acceleration energy at the UHECR sources, the spectral index at the sources and the cosmic-ray composition at the sources. For the CTG model that fits both the Auger spectrum and composition, the expected cosmogenic neutrino and gamma-ray spectra are not in tension with current measurements. Nevertheless, assuming that most of the IGRB comes from unresolved point sources, with a few more years of data Fermi/LAT might be able to set further constraints on the cosmogenic gamma-ray flux. 

In this combined fit model there are no protons present at the highest energies. The composition data of Auger at these energies does, however, allow for a non-negligible proton fraction. The cosmogenic neutrino flux at $\sim1$~EeV can be used to constrain this proton fraction. Current and future neutrino experiments with sensitivities in the range of $\sim 10^{-8}$ -- $10^{-10}$ GeV cm$^{-2}$ s$^{-1}$ sr$^{-1}$ for the single-flavor neutrino flux at $\sim1$~EeV will be able to do this for realistic source evolution models.

\acknowledgments

AvV acknowledges financial support from the NWO Astroparticle Physics grant WARP.

\end{document}